\documentclass{sig-alternate-10pt}

\usepackage{array}
\usepackage{url}
\usepackage{amsmath}
\usepackage{tikz,pgfplots}
\usetikzlibrary{plotmarks,positioning,shapes,arrows,backgrounds}

\usepackage[labelfont=bf,font={bf,small}]{caption}
\usepackage{subfigure}
\usepackage{epstopdf}
\usepackage{etoolbox}

\newlength\figureheight 
\newlength\figurewidth 

\usepackage{color,soul}

\makeatletter
\patchcmd{\maketitle}{\@copyrightspace}{}{}{}
\makeatother
\usepackage{indentfirst}

\begin{document}

\title{Draining our Glass: An Energy and Heat \\Characterization of Google Glass}
\subtitle{\vspace{1mm}Technical Report 2014-3-23, Rice University}
\author{
Robert LiKamWa$^\dag$, Zhen Wang$^\dag$, Aaron Carroll$^{\dag\ddag}$, Felix Xiaozhu Lin$^\dag$,  and Lin Zhong$^\dag$
\vspace{1mm}\\
 \affaddr{$^\dag$Rice University, Houston, TX\hspace{+3cm}$^\ddag$UNSW, Australia}\\
}

\maketitle

\graphicspath{{figures/}}

\makeatletter{}\abstract
The Google Glass is a mobile device designed to be worn as eyeglasses. This form factor enables new usage possibilities, such as hands-free video chats and instant web search. However, its shape also hampers its potential: (1) battery size, and therefore lifetime, is limited by a need for the device to be lightweight, and (2) high-power processing leads to significant heat, which should be limited, due to the Glass' compact form factor and close proximity to the user's skin. We use the Glass in a case study of the power and thermal characteristics of optical head-mounted display devices. We share insights and implications to limit power consumption to increase the safety and utility of head-mounted devices.
 
\makeatletter{}
\section{Introduction}
\begin{figure*}[ht!] 
\centering
\begin{minipage}{0.45\textwidth}
\centering
\includegraphics[width=.8\textwidth]{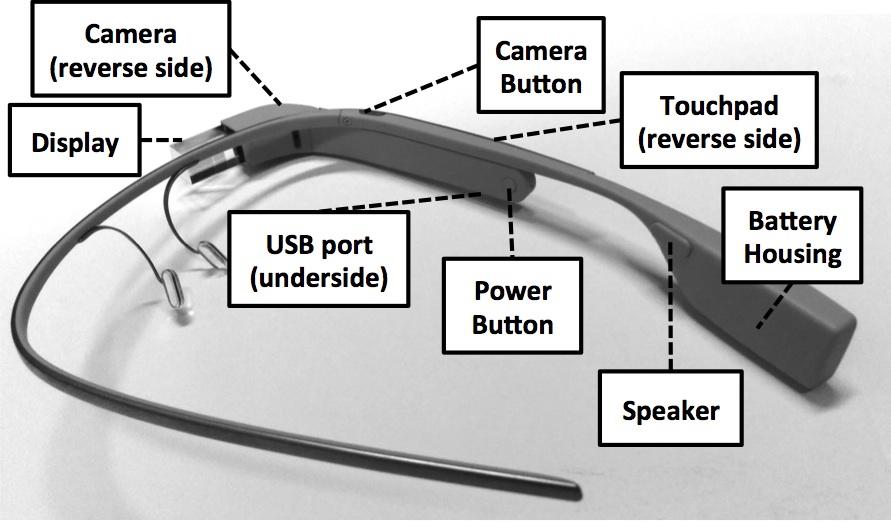}
\caption{Google Glass user interface hardware}
\label{fig:glass}
\end{minipage}
\hspace{+8mm}
\begin{minipage}{0.45\textwidth}
\centering
\includegraphics[width=.8\textwidth]{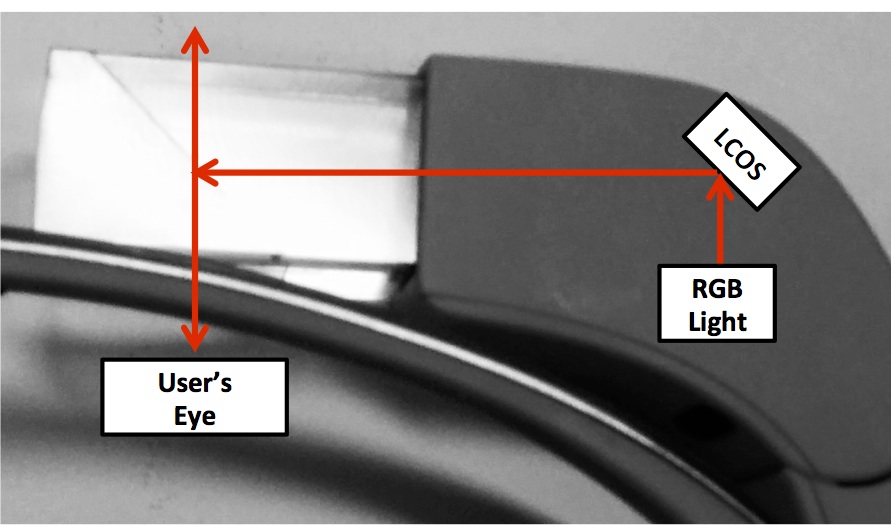}
\caption{Google Glass display projection path}
\label{fig:projection}
\end{minipage}
\end{figure*}

Optical Head Mounted Display (OHMD) devices, including~\cite{moverio, vuzix}, provide users with a hands-free display, with rich user-centric experiences and immediate access to computing resources. 
Recently, interest has drawn to Google's spectacle-shaped device, called \emph{Glass}, marking a commercial advancement in wearables, shown in Fig.~\ref{fig:glass}. 

In this article, we share results from characterizing Glass' power draw. 
Others have documented other aspects, e.g., technical specifications~\cite{googleglassteardown} and privacy, security, and social concerns~\cite{hong2013privacy,stevens2013engadget}. 
In contrast, we study Glass as an OHMD system, especially the power draw of its components and the form factor's implication on app usage and system design.

While it is tempting to treat the Glass architecture as a smartphone or tablet in a different form factor with new use cases, OHMD physical limitations magnify the value of efficiency, as compactness limits battery capacity. Moreover, contact with a user's skin will make heat generation from power draw uncomfortable and potentially dangerous. 

Thus, as low-power constraints pose the greatest technical challenge to OHMDs, we perform component-driven and usage-driven power analysis. We find that many scenarios, including mobile vision and long-term video chats, are not possible under the system's power draw. Following our analysis, we discuss hardware and software insights regarding heat constraints, display efficiency, and processor power that motivate study into efficient OHMD system design.

\makeatletter{}
\section{Glass System Overview}

Google's Glass system resembles smartphone architectures with a few notable differences: there is no cellular modem; the display is much smaller; the touchpad is distinct from the display; and a bone-conduction speaker provides audio output. 
Many other components can be found on smartphones. 

The design centers around an OMAP4430 system-on-chip (SoC), which includes a dual-core ARM Cortex-A9 as the main CPU, a dual-core ARM Cortex-M3, an SGX540 GPU, a Display Subsystem (DSS), an Image and Video Accelerator (IVA) and a DSP~\cite{omap4430}. 
The OMAP4430 is also used in the Motorola Droid RAZR, the LG Optimus 3D Max, the Samsung Galaxy Tab 2, and many other mobile devices.

Because of the OHMD form factor, the OLED or LCD smartphone display is absent, replaced with a Liquid-Crystal-on-Silicon (LCOS) projection display, shown in Fig.~\ref{fig:projection}. 
As an LED is filtered to shine red, green, or blue light onto the LCOS, the display shows the respective color component contribution. The image is projected on a semi-reflective mirror directly in front of the user's right eye. This allows content to be seen not only by the user, but also from the front of Glass. As red, green, and blue images are flashed rapidly, the user perceives a full-color image. According to \cite{brennesholtz2008projection}, component images must cycle at 540 Hz for color video.

Glass employs a unique bone-conduction speaker to allow the user to hear clearly while minimizing the sound to people not using the device. 
A Synaptics touchpad component provides touch functionality on the side of the Glass.

A 3.7 volt LiPo battery with a capacity of 2.1~Wh provides power to the Glass. The battery sits behind the ear, counterbalancing the rest of the device over the user's ear. 

Glass runs Android OS, v. 4.04 ``Ice Cream Sandwich''.

\makeatletter{}
\section{Use Cases}\label{sec:usage}
As a hands-free device with display, camera, and radios, Glass creates new usage potential. 
While not an exhaustive list, we discuss potentially useful OHMD apps. Most are advertised by Google and are marked with *~\cite{howitfeels}.

\textbf{Real-time Hands-free Notifications*:} 
Users can field phone calls, text messages, e-mails, etc., without reaching into a pocket or bag.  This mitigates user disruptions, e.g., a user can ride a bike while addressing a text message.

\textbf{Hands-free visual and audio instructions*:}
OHMDs can access instructions and materials related to a user's activity. This is useful for a cook following a recipe, an artist consulting reference material, or a tourist asking for language translation. OHMDs could even assist surgeons using CT or X-ray images during surgery~\cite{glass4surgery}.

\textbf{Instant Connectivity Access*:} 
Users can instantly access network-based information, including email inboxes, web searches, local news, stock market tickers, home security camera feeds, or social networks. 

 \textbf{Instant Photography/Videography*:} 
OHMD cameras allow users to take pictures and video clips without pointing a smartphone. This is useful for journaling momentary experiences or recording a user's perspective of a scene. Video can also be streamed in real-time for engaging video chats.

\textbf{Augmented Reality*:} 
In augmented reality (AR), virtual objects are displayed as part of the physical environment. 
AR can overlay path directions, assist in interior design, or enable other immersive visualization apps. The Glass Developer documentation lists ideas for simple AR games which use voice and inertial motion sensors for interaction~\cite{googleARgames}.

\textbf{Continuous Mobile Vision:} 
A technology that many expect out of OHMDs is the ability to observe and understand a scene through computer vision. 
Face detection and recognition are feasible, while discouraged by Google. Vision can be used for many other tasks, including text recognition, geometric scene understanding, and contextual life logging.

These scenarios would provide a rich experience to the OHMD user. 
We benchmark representative workloads in Section~\ref{sec:bench}. 
These potential use cases dictate device requirements and thus influence the system architectural design.

\makeatletter{}\begin{figure}[t!] 
\centering
\makeatletter{}\begin{tikzpicture}[font=\small]
\begin{axis}[xlabel={Processor Frequency (MHz)}, ylabel={Power Consumption (mW)}, width=3in, height=2.6in,ymin=0,ymax=2200, xmin=201, xmax=900,
legend style={at={(1.1,0.015)},anchor=south east,legend cell align=left}
]
\addplot[scatter, color=orange,
scatter/use mapped color={draw=orange, scale=1}]
	      table[x=x,y=y] {
    x        y    
300	1306.9
600	1717.4
800	2090.2
};
\addplot[scatter, mark=square*, color=red,
scatter/use mapped color={draw=red,scale=0.8}]
	table[x=x,y=y] {
    x        y  
300	1191.44
600	1480.1
800	1690.3
};
\addplot[scatter, mark=triangle*, color=black!60!green,
scatter/use mapped color={draw=black!60!green,scale=1.2}]
      table[x=x,y=y] {
    x        y     
300	546.64
600	1149.1
800	1547.9
};
\addplot[scatter, mark=diamond*, color=blue,
scatter/use mapped color={draw=blue,scale=1.2}]
      table[x=x,y=y] {
    x        y     
300	421.83	
600	896.3	
800	1132.4	
};

\addlegendentry{2 core, Screen on}
\addlegendentry{1 core, Screen on}
\addlegendentry{2 core, Screen off}
\addlegendentry{1 core, Screen off}
\end{axis}
\end{tikzpicture}
 
\caption{Power consumption vs.\ core use, with one core and two cores at 100\% CPU Utilization, with screen on and screen off}
\label{fig:coreuse}
\end{figure}
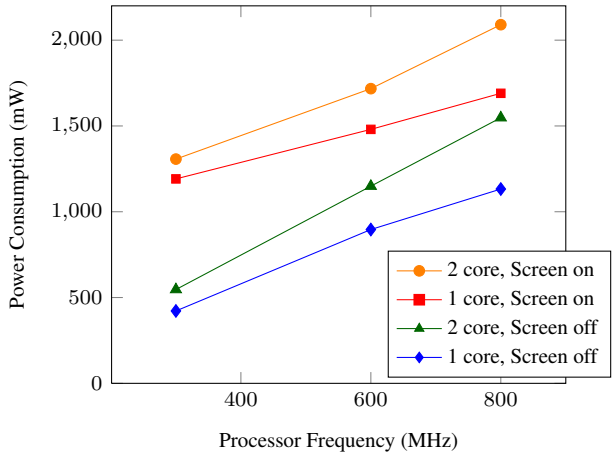
\newpage
\section{Power Measurement}\label{sec:bench}
We benchmark the power draw and CPU usage. To measure power consumption, we use a Monsoon Power Monitor~\cite{monsoon} in place of the Glass battery. 
The monitor provides 4\,V and records power draw. For CPU utilization, we use the command \texttt{top}, which has a 5--7\% CPU utilization overhead.

\subsection{Power by component}
We first explore how a component's use contributes to overall system power. We configure and utilize each component  while keeping the rest of the system constant. Our measurements thus reveal the \emph{rise in system power consumption} while a component is used.

\textbf{OMAP4430 SoC:}
The OMAP is a major contributor to Glass' power consumption. 
We check the status of its modules by using the \texttt{omapconf}~\cite{omapconf} diagnostic tool.
By default, many modules are disabled even when the screen is on, including the Cortex-M3, GPU, IVA and DSP. This leaves the Cortex-A9 as the major active computational component.

Unlike smartphones, the Glass uses the DSS instead of the GPU to perform surface composition and to stream frames to the display. Thus, the DSS is on while the screen is on, and the GPU is disabled by default. 
The GPU may be invoked by software, e.g., for OpenGL applications.
The IVA is activated for encoding/decoding video recording and playback.

The main CPU (the Cortex-A9) can be set to four frequencies: 300 MHz, 600 MHz, 800 MHz, and 1 GHz. Raised frequencies increase performance, but draw more power. At high temperatures, Glass firmware limits the frequency to 600 MHz or 300 MHz to cool down by reducing power.
We run shell scripts on one Cortex-A9 core and both cores at 100\% CPU utilization with the screen on and off. We then measure the power draw of the Glass system, shown in Fig.~\ref{fig:coreuse}. While we can briefly set the Glass to 1 GHz, the system rigorously decreases the frequency to reduce heat, prohibiting us from taking robust 1 GHz measurements.

\begin{figure}[t!] 
\centering
\begin{minipage}[b]{0.48\linewidth}
\centering
\makeatletter{}\begin{tikzpicture}[font=\small]
\begin{axis}[xlabel={Screen Brightness}, ylabel={System Power (mW)}, width=1.5in, height=1.4in]
\addplot [color=red, domain=0:255, mark=none] {1005.5+0.7961*\x};
\addplot[scatter, only marks, color=blue,
	scatter/use mapped color={draw=blue,scale=0.5}]
      table[x=x,y=y] {
    x        y     class
25	1031	0
50	1040	0
75	1071	0
100	1085	0
125	1103	0
150	1117	0
175	1144	0
200	1171	0
225	1186	0
250	1204	0
};
\end{axis}
\end{tikzpicture}
 
\end{minipage}
\hspace{0.1mm}
\begin{minipage}[b]{0.48\linewidth}
\centering
\makeatletter{}\begin{tikzpicture}[font=\small]
\begin{axis}[xlabel={Speaker Volume}, ylabel={Speaker Power (mW)}, width=1.5in, height=1.4in]

	\addplot[scatter, only marks, color=blue,
	scatter/use mapped color={draw=blue,scale=0.5}]
      table[x=x,y=y] {
    x        y     class
100	407.747669100001	0
95	402.448972199998	0
90	418.376150600001	0
85	406.1704481	0
80	411.2003951	0
75	396.294025400001	0
70	388.5854247	0
65	416.190363699999	0
60	397.1939988	0
55	439.4441042	0
50	406.8771861	0
45	405.6430447	0
40	409.7692304	0
35	408.3011913	0
30	404.3872617	0
25	382.588384300001	0
20	325.405928900002	0
15	250.240168600001	0
10	175.536364299998	0
5	100.160523	0
};
\end{axis}
\end{tikzpicture} 
\end{minipage}
\caption{(Left) System power draw vs.\ screen brightness. (Right) Speaker power draw vs.\ speaker volume.}
\label{fig:brightness}
\end{figure}
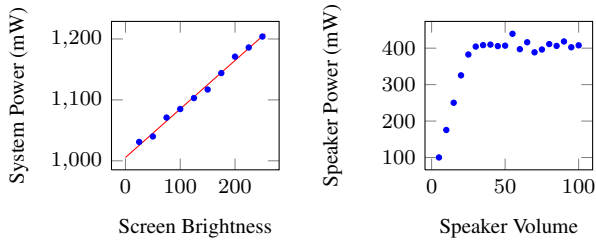

\textbf{Screen: }
Glass sets the screen brightness on a 25--255 scale depending on the sensed
ambient brightness. We set the brightness by writing to a device file while using
a static app with static screen content. As shown in Fig.~\ref{fig:brightness}, the brightness affects the Glass' power consumption.  The screen content, including its colors, does not affect the power draw. This is similar to LCDs but in contrast to OLED displays. 

We measure that Glass draws 1028 mW when the screen is at a brightness of 25. Glass draws 1204 mW at a brightness of 255. By contrast, with the screen off, but the system active, the Glass draws 334 mW. Thus, when using the screen, the system draws a static power of 674 mW and dynamically draws another 196 mW depending on the brightness level. 
\begin{equation*}
P_\mathrm{screen} = 674\,\mathrm{mW} + 196\,\mathrm{mW}\times(\mathrm{brightness}/255)
\end{equation*}

The high static power draw of 674 mW is likely due to the activation of the DSS display subsystem, its rendering of the screen content, and the transmission to the LCOS.

As the display is close to the eye, we expected the dynamic power to be orders of magnitude lower than that of a smartphone. Display power is typically proportional to $D^2$ where $D$ is the distance from the screen to the eyes, if all other factors remain the same~\cite{zhong2005mobisys}. 
Since $D$ for the Glass is over 10$\times$ smaller than that for a phone, we
expected a display draw of < 5 mW, as the iPhone 4 LCD consumes
$\sim$420 mW~\cite{iphone4lcd} at full brightness. The actual projection consumes up to 196 mW, much higher than expected. This is likely due to luminance drops in the display path. First, the color filter reduces the luminance of the LED projection. Luminance is further reduced by $\sim$40\% in reflection off of typical LCOS devices \cite{holoeye}. Finally, the semi-reflective mirror in front of the user's eye incurs a >50\% drop in the reflective optics. These drops necessitate high-power projection at the source. 

We perform the rest of our measurements with the screen brightness fixed to 25, suitable for our office environment.

\textbf{Bone-Conduction Speaker: }
Using the bone-conduction speaker consumes $\sim$410 mW when the volume is at or above 35\%. The sound production is not louder above 35\%. Below 35\%, power consumption decreases, down to 100 mW at 5\%. 
When playing audio through a USB Earpiece, Glass draws 18--30 mW.
For comparison, we measure that a Galaxy Nexus draws $\sim$200 mW when
using its speakers.

\textbf{Inertial Motion Unit: }
The Android API can sample the accelerometer and gyroscope at either 100 Hz or 200 Hz. 
Sampling consumes $\sim$29 mW, regardless of sampling speed.

\textbf{Audio Recording: }
Using the microphone on the Glass to record audio consumes an additional 96 mW.

\begin{table}[t]
\centering
\caption{Glass power draw in different usage scenarios}
\small
\begin{tabular}{|l|l|l|}
\hline
\textbf{Usage Case} & \textbf{Power Consumption} & \textbf{Battery Life}\\
\hline
Idle & 22 mW & 95 hours\\
\hline
System active, screen off &334 mW & 377 min. \\
\hline
Static timeline card& 1030 mW & 122 min.\\
\hline
Timeline swiping & 1315 mW & 96 min.\\
\hline
"Ok Glass" card & 1204 mW & 105 min.\\
\hline
Main menu card & 2361 mW & 53 min.\\
\hline
Internet page load & 2009 mW (3 sec.) &  1260 page loads\\
\hline
Web page viewing & 1171 mW & 107 min.\\
\hline
Web page scrolling & 1505 mW & 84 min.\\
\hline 
Phone calls & 1257 mW & 100 min.\\
\hline 
Text message & 1387 mW (1.3 sec.) & 4200 messages\\
\hline 
Image capture & 2927 mW (3.3 sec.) & 782 images\\
\hline 
Video capture & 2963 mW & 43 min. \\
\hline 
Video chat & 2960 mW & 43 min.\\
\hline 
Static application & 1023 mW & 123 min.\\
\hline 
Camera preview callback & 2366 mW & 53 min.\\
\hline 
OpenCV face detection & 3318 mW & 38 min.\\
\hline

\end{tabular}
\label{table:scenario}
\end{table}

\textbf{WiFi/Bluetooth: }
We measure the Glass before activating WiFi/Bluetooth and during a file download. 
While downloading at 538 kbps over Bluetooth, Glass draws an additional 743 mW. 
On WiFi, at 734 kbps, Glass draws 653 mW.

\subsection{Power by Usage Scenario}

We next collect average power measurements as we place the Glass under various app workloads, as shown in Table~\ref{table:scenario}.

\textbf{Idle Power: }
Glass exhibits an efficient idle mode, using background
processing to sense wake-up events, such as notifications or accelerometer gestures. Processors, radios, and the display are held in an inactive low-power mode. The idle Glass consumes 22.3 mW for a lifetime of over 90 hours.

When the system wakes up, the CPU briefly rises to 1 GHz to quickly wake the system up, and subsequently returns to 300 MHz, waiting for user interaction.
This wake-up process takes 1.9 seconds and consumes 1398 mW on average.

\textbf{Menu Navigation: }
After waking up, Glass opens a heads-up menu. Primary interaction is through the touchpad: the user swipes to highlight items, and taps to activate one. This lets the user access instant connectivity, address notifications, or initiate hands-free visual and audio interactions.

When the user is observing a static timeline card, the CPU utilization is near 0\% and the Glass draws $\sim$1030~mW, giving the Glass a battery lifetime of 2 hours in this scenario. 
When the user is swiping through the cards, a "mediaserver" process is invoked
requiring approximately 8--10\% CPU utilization at 600 or 800 MHz, to load the new processes. The system draws additional power to a total of 1315 mW while being swiped, but this only lasts as long as the swipe persists. 

Voice command is available on the main menu card, where the Glass displays the time.
The system waits for a user voice input and consumes 1204 mW, which is about 200 mW more than static timeline cards due to the background voice recognition process with microphone sampling.
When the keyphrase "OK Glass" is spoken, a submenu of speakable items appears. 
A user selects an item for launch by using the touchpad or reading its title. 
Glass draws 2361 mW while the user is navigates the menu by voice.
Actively recognizing voice uses 30\% CPU utilization at 300 MHz for the voice recognition process, and samples the microphone sensor.

\newpage
\textbf{Internet Browsing: }
The built-in internet browser application does not have an address bar to enter a URL, but a user can perform a Google voice search and access the web page result. Once in a webpage, the user places two fingers on the touchpad and moves their head to pan a website.

Loading a Wikipedia page takes $\sim$3 seconds, incurs CPU utilization of 50\%, and draws 2009 mW. Viewing the loaded page draws 1171 mW, while scrolling the page draws 1505 mW.
This is comparable to browsing on a Galaxy S III~\cite{carroll2013apsys}.
As with the smartphone, the high power draw is mainly due to the high static power of using the display. 

\textbf{Telephony: }
Glass can operate as a headset, routing phone call audio between the phone and the Glass for hands-free communication. This uses the microphone, speaker, and the Bluetooth card. 
Running the telephony incurs a 35\% CPU utilization and draws 1257 mW, for a battery life of 102 minutes. Receiving a text message draws 1387 mW for 1.3 seconds. 4200 messages can be received on a single charge.

This is disproportionately high; a Bluetooth headset consumes an order of magnitude less power. This is likely because the entire Glass system is awake and using the high-power Cortex-A9 to perform simple telephony tasks.

\textbf{Image/Video Capture and Streaming: }
Glass can instantly capture photos and videos on demand. Unfortunately, the image capture and processing draw significant power. 
Using the built-in camera app to take a picture consumes a peak of 4629 mW and an average of 2927 mW for 3.3 seconds. Glass takes fewer than 800 pictures on a single charge. 

For video recording, our measurements show that the Glass draws 2963 mW while using the camera app.  
At that power draw, the Glass can take a video for 45 minutes on a single battery charge.
Streaming a video call over WiFi draws roughly the same amount of power, consuming an average of 2960 mW, allowing a 45 minute video cat on a single charge. 

\textbf{Vision Application Usage: }
Developers may write applications using the standard Android SDK and deploy them on the Google Glass.  Running a static application draws similar power to that of a static timeline card: 1023 mW. In this scenario, the system uses around 18\% CPU utilization.

Glass vision apps can be developed by using the preview frame callback of the Android camera service. This is a significant workload for the Glass; the act of running the camera service and previewing a frame uses 85\% CPU utilization at 600 MHz. Furthermore, the IVA and GPU are activated. As a result, the system draws a total power of 2366 mW.

To operate on the frame, we use the OpenCV library with Android bindings. Running a Face Detection app brings the device to a full 100\% CPU Utilization at 600 MHz. This draws a large 3318 mW, for only 38 minutes of battery life.
 
\makeatletter{}

\section{Temperature Characterization}\label{sec:temp}
As with all electronics, power is linked to temperature, as nearly all power is dissipated as heat. Because Glass makes contact with its user's face, heat is a critical safety issue. 
\subsubsection*{Measurement}

To characterize thermal behavior, we use an ST-380 surface thermometer pointed where Glass makes contact with the user's temple, the face region behind the eyes. This tends to be the warmest part of Glass, as it is where the OMAP is located. The OMAP is the major element that contributes to the temperature in the measured area; changing screen brightness or using the speaker did not change the reported temperature. 

Unlike our power measurements, we used a Glass that we did not tamper with.
The room was held at 23$^{\circ}$C with adequate ventilation, and the Glass is 31.1$^{\circ}$C while charging. 
After disconnecting the charger, we started a video chat.
The thermometer reported a rapid rise, shown in Fig.~\ref{fig:thermal}. 
After 120 seconds, the Glass rose to 39$^{\circ}$C. 
The rise eventually slowed, reaching a stable 51.9$^{\circ}$C (124$^{\circ}$F).
We repeated the process with a static idle app, reaching a stable 35.2$^{\circ}$C.

\subsubsection*{Modeling}

We next use Newton's Cooling Law~\cite{louis1993convective} to relate SoC power draw to the steady-state temperature difference ($\Delta T$) of the Glass with its environment. We model  $\Delta T$ as proportional to the power dissipation ($P$) and inversely proportional to the surface area ($A$) and the convection coefficient~($h$). The initial device temperature does not affect $\Delta T$. 
\begin{equation*}
P = dQ/dt = hA\Delta T
\end{equation*}

Newton's Cooling Law also models how temperature changes over time to reach the steady-state temperature. For a device of heat capacity $C$, the temperature at a given time is:
\begin{equation*}
T(t) =T(0) + (1-e^{-(C/(hA))t})\Delta T
\end{equation*}

Thus, power consumption of the device dictates the steady-state temperature difference ($\Delta T$);  how fast the temperature rises is determined by the mechanical and material properties of the device and its environment, i.e., $C$, $h$, and $A$. 

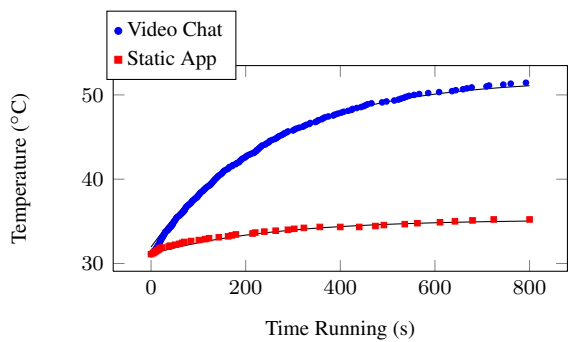
\begin{figure}[!t] 
\centering
\makeatletter{}\tikzset{mark size=1.3}
\begin{tikzpicture}[font=\small]
\begin{axis}[xlabel={Time Running (s)}, ylabel={Temperature ($^{\circ}$C)}, width=3in, height=1.7in,
legend style={at={(-0.015,0.93)},anchor=south west,legend cell align=left}
]
\addplot[scatter, only marks, 
	range=-250:250, color=blue,
	scatter/use mapped color={draw=blue, fill=blue, scale=0.8},
	scatter src=\thisrow{class}]
      table[x=x,y=y] {
    x        y     class
0	31.11111111	0
3.103952	31.22222222	0
3.823836	31.33333333	0
5.35969	31.44444444	0
7.35969	31.55555556	0
9.567408	31.66666667	0
10.703358	31.77777778	0
11.759526	31.88888889	0
13.07158	32	0
14.431686	32.11111111	0
15.759718	32.22222222	0
17.0393	32.33333333	0
17.759706	32.44444444	0
18.799728	32.55555556	0
19.839132	32.66666667	0
20.65552	32.77777778	0
21.647508	32.88888889	0
23.216542	33	0
24.240978	33.11111111	0
25.343244	33.22222222	0
26.52748	33.33333333	0
27.935674	33.44444444	0
28.751272	33.55555556	0
29.711526	33.66666667	0
30.462982	33.77777778	0
34.478748	33.88888889	0
36.190888	34	0
37.872294	34.11111111	0
39.29534	34.22222222	0
41.663016	34.33333333	0
42.927062	34.44444444	0
44.975366	34.55555556	0
45.695104	34.66666667	0
46.799118	34.77777778	0
47.710676	34.88888889	0
48.686988	35	0
50.831174	35.11111111	0
51.822702	35.22222222	0
52.623026	35.33333333	0
54.207062	35.44444444	0
56.367022	35.55555556	0
60.126982	35.66666667	0
61.375364	35.77777778	0
63.886824	35.88888889	0
66.062522	36	0
68.014744	36.11111111	0
69.806732	36.22222222	0
71.10394	36.33333333	0
71.711908	36.44444444	0
72.79888	36.55555556	0
74.062592	36.66666667	0
75.326444	36.77777778	0
78.398698	36.88888889	0
80.206742	37	0
83.134762	37.11111111	0
84.878534	37.22222222	0
86.062592	37.33333333	0
87.614358	37.44444444	0
90.558258	37.55555556	0
92.670538	37.66666667	0
95.98257	37.77777778	0
97.742458	37.88888889	0
99.326972	38	0
100.60644	38.11111111	0
102.302774	38.22222222	0
103.614646	38.33333333	0
105.42244	38.44444444	0
108.621948	38.55555556	0
110.766238	38.66666667	0
112.062548	38.77777778	0
114.718236	38.88888889	0
116.301954	39	0
119.328994	39.11111111	0
122.91012	39.22222222	0
125.534246	39.33333333	0
127.05424	39.44444444	0
128.75013	39.55555556	0
130.41376	39.66666667	0
132.14298	39.77777778	0
133.693818	39.88888889	0
135.390806	40	0
137.518126	40.11111111	0
139.838876	40.22222222	0
141.357798	40.33333333	0
143.917624	40.44444444	0
146.654758	40.55555556	0
149.645474	40.66666667	0
150.941828	40.77777778	0
152.733458	40.88888889	0
154.526184	41	0
158.349854	41.11111111	0
162.493716	41.22222222	0
164.158626	41.33333333	0
165.629374	41.44444444	0
171.325454	41.55555556	0
172.781352	41.66666667	0
178.685066	41.77777778	0
180.829084	41.88888889	0
185.56549	42	0
187.244842	42.11111111	0
188.895096	42.22222222	0
191.405142	42.33333333	0
193.32513	42.44444444	0
198.108776	42.55555556	0
200.174316	42.66666667	0
202.540882	42.77777778	0
206.797604	42.88888889	0
211.501118	43	0
217.51633	43.11111111	0
219.549322	43.22222222	0
222.780876	43.33333333	0
222.988516	43.44444444	0
226.621576	43.55555556	0
228.444518	43.66666667	0
230.572508	43.77777778	0
232.684458	43.88888889	0
235.036346	44	0
241.308466	44.11111111	0
243.36013	44.22222222	0
247.676618	44.33333333	0
249.788246	44.44444444	0
251.356026	44.55555556	0
256.508112	44.66666667	0
262.156052	44.77777778	0
265.564404	44.88888889	0
267.90032	45	0
270.829728	45.11111111	0
275.644424	45.22222222	0
279.515936	45.33333333	0
281.755722	45.44444444	0
286.571564	45.55555556	0
289.388152	45.66666667	0
300.091966	45.77777778	0
303.163982	45.88888889	0
307.163968	46	0
313.003478	46.11111111	0
320.811838	46.22222222	0
323.803458	46.33333333	0
327.931886	46.44444444	0
334.509056	46.55555556	0
336.570992	46.66666667	0
347.322844	46.77777778	0
351.370898	46.88888889	0
353.675736	47	0
360.618682	47.11111111	0
363.706808	47.22222222	0
367.372084	47.33333333	0
374.698988	47.44444444	0
385.211656	47.55555556	0
388.746424	47.66666667	0
394.842568	47.77777778	0
401.81809	47.88888889	0
407.227758	48	0
416.633794	48.11111111	0
423.721808	48.22222222	0
425.930274	48.33333333	0
434.745934	48.44444444	0
443.833704	48.55555556	0
449.689446	48.66666667	0
453.514758	48.77777778	0
457.705918	48.88888889	0
465.849554	49	0
488.3139	49.11111111	0
497.593304	49.22222222	0
513.976694	49.33333333	0
522.745976	49.44444444	0
528.39265	49.55555556	0
534.53631	49.66666667	0
539.497402	49.77777778	0
545.38439	49.88888889	0
555.831912	50	0
566.761346	50.11111111	0
585.783762	50.22222222	0
608.823444	50.33333333	0
636.087616	50.44444444	0
644.13512	50.55555556	0
659.191698	50.66666667	0
667.574952	50.77777778	0
677.68654	50.88888889	0
708.982184	51	0
714.390428	51.11111111	0
744.437708	51.22222222	0
760.18271	51.33333333	0
792.374378	51.44444444	0
};
\addplot[scatter, only marks, mark=square*, 
	range=-250:250, color=red,
	scatter/use mapped color={draw=red, fill=red, scale=0.8},
	scatter src=\thisrow{class}]
      table[x=x,y=y] {
          x        y     class
0	31.11111111	1
4.576084	31.22222222	1
7.232932	31.33333333	1
10.402	31.44444444	1
13.376148	31.55555556	1
16.128196	31.66666667	1
17.53582	31.77777778	1
22.40002	31.88888889	1
37.119588	32	1
40.130392	32.11111111	1
50.97564	32.22222222	1
59.90364	32.33333333	1
65.824356	32.44444444	1
70.367876	32.55555556	1
84.6385	32.66666667	1
101.151408	32.77777778	1
114.469076	32.88888889	1
122.534712	33	1
143.166224	33.11111111	1
162.974068	33.22222222	1
170.977292	33.33333333	1
179.230088	33.44444444	1
214.55776	33.55555556	1
220.221888	33.66666667	1
239.51762	33.77777778	1
263.32518	33.88888889	1
289.307892	34	1
302.779824	34.11111111	1
322.972516	34.22222222	1
356.635696	34.33333333	1
400	34.33333333	1
440	34.33333333	1
474.265336	34.44444444	1
492.189144	34.55555556	1
536.21646	34.66666667	1
562.96982	34.77777778	1
610.714304	34.88888889	1
642.199512	35	1
679.448768	35.11111111	1
724.249376	35.22222222	1
800	35.22222222	1
    };
\addplot [color=black, domain=0:800, mark=none]{27.9*(1-exp(-0.004*(x+84)))+24};
\addplot [color=black, domain=0:800, mark=none]{11.2*(1-exp(-0.004*(x+252)))+24};
\addlegendentry{Video Chat}
\addlegendentry{Static App}
\end{axis}

\end{tikzpicture} 
\caption{Temperature samples vs. time running an application}
\label{fig:thermal}
\end{figure}

We fit the model to our data in Fig.~\ref{fig:thermal}. While the OMAP consumes 3~watts during video chat, we model that $\Delta T\,=\,28.9^{\circ}C$, and $C/hA\,=\,0.0040\,s^{-1}$. For 1~watt draw in static app usage, we model that $\Delta T~=~11.2^{\circ}C$, and $C/hA\,=\,0.0040\,s^{-1}$. Thus, we have the following observations for Glass when used in a similar office environment to ours:
\begin{itemize}
\item  For every watt drawn by the OMAP, the steady-state surface temperature will rise by $\Delta T\approx10^{\circ} \mathrm{C}$ higher than its environment. 

\item The Glass surface temperature will rise by 90\% of $\Delta T$ in 10 minutes.

\end{itemize}

That is, the temperature difference from the environment is roughly proportional to the SoC's power draw, while the rate of heating is dependent only on its environment.

Thus, \emph{long-term average power consumption of a process dictates stable temperature and should be constrained for heat safety}. This may incur energy inefficiency, as the time it takes to run a low-power process may exceed the optimal point to minimize energy use. However, constraining average power will keep the device temperature in safe limits.

We collect measurements while the Glass is not worn by a user. 
A user may raise the Glass' temperature due to the user's body being warmer than that of the environment. 
Indeed, the authors experience that the device feels hot while performing intensive tasks. 
The Glass could also experience an even higher temperature with a harsh environment, e.g., in direct sunlight in 38$^{\circ}$C weather.

\subsubsection*{Health Implications}

High temperatures put pressure on the human body to regulate its temperature to below 37$^{\circ}$C, which it does by dilating blood vessels, increasing heart rate, raising skin temperature, and activating sweat glands, as explained in \cite{k1994ergonomics}. These reactions lead to reduced comfort and potential cardiovascular problems. Indeed, blood vessel damage can occur with continuous contact to surfaces at temperatures as low as 38--48$^{\circ}$C
, leading to permanent skin damage, such as \emph{erythema ab igne}~\cite{riahi2012laptop}. Thus, while the measured surface temperatures are common for smartphones and tablets, the device surface temperatures of 50+$^{\circ}$C are not well-suited for a head-mounted device with large durations of skin contact.

\makeatletter{}\section{Insights and Recommendations}
Glass consumes an amount of energy close to that of a smartphone. This, paired with skin contact with the user, places thermal issues as a first-class design constraint. The high power draw also limits usage time due to the small battery size. Among the use cases in Section~\ref{sec:usage}, \emph{only hands-free notifications and instant connectivity are safe and feasible}.

Long-term visuals and audio are constrained by the power draw of using the display and speaker. Photography and video chats are further limited; camera usage draws $\sim$3\,W, heating Glass $28^{\circ}\mathrm{C}$ above its environment. While the form factor suggests easy image capture and vision opportunities, power draw limits the longevity and safety of such apps. 

Thus, the Glass design is not suitable for always-on use, including long-term AR and mobile vision. While off-loading perception tasks to the cloud is difficult, due to transmission cost and need for connectivity~\cite{han2013case}, onloading computations is impractical due to high power draw. For these tasks, a smartphone-like system cannot achieve the required efficiency. Instead, we propose the following considerations.

\subsubsection*{Display Efficiency}
In addition to computational power expense, the system consumes an additional 675 mW to show screen content, and up to another 200 mW depending on the brightness. This prohibits long-term use of the screen. We recommend an adoption of efficient content generation and projection. 

\textbf{Static Display Subsystem:}
While Glass should be able to display active scenes, the DSS consumes too much for slow-changing content, e.g., displaying a cookbook. One potential solution could be to use a hybrid DSS with a bistable mode to provide static content for low-power viewing. 

A hybrid DSS would introduce system research challenges of optimally deciding which mode to use, as well as supporting migration between active and static DSS modes.  
 
\textbf{Efficient Projection:}
The LCOS projection consumes two orders of magnitude more power than its proximity requires, due to luminance drops in the display path. Replacing this projection with a transparent OLED screen in front of the eye would reduce power draw by avoiding luminance drops. 
Moreover, the see-through nature of OHMDs means that for many apps, few
pixels are active at any time. This is optimal for OLED, as inactive pixels do not draw power. 

\subsubsection*{Computational Efficiency}
\indent Computational power draw is the major contributor to Glass' high power consumption. This strongly motivates principles of energy-proportionality and
optimizing the common case.

\textbf{Heterogeneous Computing:}
OHMDs present conflicting processor requirements: always-on cases demand high
efficiency at low throughput, medium-term power is limited by battery life
and device heat, and user I/O demands intermittent high throughput to limit
visible latency. While the Cortex-A9 provides high throughput, merely onlining the
CPU consumes $\sim$330\,mW, making it unsuitable for always-on cases. Core
heterogeneity is a known solution to this problem. We believe this is not
merely convenient for longer battery life, but in fact required for the OHMD form
factor to deliver on its promised functionality. System support for core
heterogeneity is an active area of research~\cite{lin2012asplos, lin2014asplos}, and our results
provide a further data point in motivating this work.

\textbf{Vision Acceleration:}
Vision processing on the CPU draws a high 3.3 watts on the Glass, limiting the use
cases for readily-available camera capture. For these apps to be feasible,
an efficient vision processing unit, such as the Movidius Myriad \cite{movidius},
will be needed. Existing application-specific circuity, such as GPUs and video
codecs, have well-defined interfaces (OpenGL, MPEG, etc.). We see a research
opportunity in exposing vision acceleration to user apps,
including issues of resource management, multiplexing, isolation,
etc.

\subsubsection*{Responsible Thermal Control}
Because of user contact, heat is a paramount concern, as every watt generates a 10$^{\circ}$C rise. While lowering power draw is necessary to reduce heat, optimizing heat dissipation and introducing system policies can alleviate thermal issues.

\textbf{Improved Heat Dissipation:}
Heat dissipation can be partially addressed with physical redesign. To increase user comfort and safety, the Glass form factor should distance processing elements from skin contact and use heat sinks to dissipate OMAP heat over a wider surface area.

\textbf{Thermal Regulation:}
Thermal regulation is well-studied, usually through limiting
power draw to the thermal design point (TDP). We have shown that power should be constrained to a safe user limit much lower than typical TDP. Moreover, Glass exceeds this limit under typical operation. Thus, the system should be designed for graceful degradation.  For instance, when the thermal limit is reached during video chat, the resolution, frame rate, encoding quality, bitrate, etc.\
\emph{must} be reduced to avoid physical harm. Typical approaches of reducing frequency, disabling cores, or killing
processes are not satisfactory, as this situation occurs in everyday
usage.

Additionally, thermal limits are not hard, i.e., peak power need not be strictly limited. As Fig.~\ref{fig:thermal} shows, the heating time constant is relatively long.
Efficient thermal regulation should control average and not peak power, but it must also be aware of the time period over which this average must be maintained. Strategies such as \cite{raghavan2013computational} can computationally sprint to briefly raise peak power while managing heat.

\subsubsection*{Low Power I/O}
The camera, screen, and speaker are designed for high power, high quality use. Glass should have I/O quality mode options for power reduction strategies.

\textbf{Scalable Imaging:}
Using the camera reduces battery life to one hour due to the high-power IVA and image sensor, regardless of capture quality. Image sensor energy cost should be proportional to the frame rate and resolution, using techniques such as ~\cite{likamwa2013mobisys}. Also, a multi-rate IVA could provide low-power processing when high resolution is not required.

\textbf{Notification LED:}
When a text message arrives, the DSS is activated, consuming precious energy. Instead, an LED on the display could notify the user, to avoid display activation. This is an effective strategy for smartphone notifications, and should be adopted for OHMD energy efficiency.

\textbf{Moving Coil Speaker:}
The bone-conduction speaker is energy expensive for continuous use, such as for music or video playback or for game audio. A supplementary moving coil speaker would allow a choice for lower-power audio.

\section{Conclusion}
We use Google Glass as a platform for studying system aspects of wearable devices that are constrained by form factor and battery life. While it has suboptimal power draw and heat dissipation, it provides an exciting public introduction to wearable devices and a base for future OHMD design. 

The Glass device motivates deep investigation into wearable systems. We find that the high performance, significant power draw, and thermal concerns present several OHMD research opportunities towards improved efficiency. 

 \small
\bibliographystyle{plain}
\bibliography{./bib/abr,./bib/glass,./bib/zhong}
\end{document}